\documentclass[cits]{PoS}
\usepackage[utf8]{inputenc}
\usepackage{slashed}

\title{Computation of disconnected contributions to nucleon observables}

\ShortTitle{Computation of disconnected contributions to nucleon observables}

\author{Constantia  Alexandrou\\
Department of Physics, University of Cyprus, P.O. Box 20537, 1678 Nicosia, Cyprus, and\\  
Computation-based Science and Technology Research Center, Cyprus Institute, 20 Kavafi Str., Nicosia 2121, Cyprus \\  
E-mail: \email{alexand@ucy.ac.cy}}
\author{Vincent Drach\\
NIC, DESY, Platanenallee 6, D-15738 Zeuthen, Germany\\
E-mail: \email{vincent.drach@desy.de}}
\author{Karl Jansen\\
NIC, DESY, Platanenallee 6, D-15738 Zeuthen, Germany\\
E-mail: \email{karl.jansen@desy.de}}
\author{Giannis Koutsou\\
Computation-based Science and Technology Research Center, Cyprus Institute, 20 Kavafi Str., Nicosia 2121, Cyprus\\
E-mail: \email{g.koutsou@cyi.ac.cy}}
\author{\speaker{Alejandro Vaquero Avilés-Casco$^{a}$} \\
Computation-based Science and Technology Research Center, Cyprus Institute, 20 Kavafi Str., Nicosia 2121, Cyprus\\
E-mail: \email{a.vaquero@cyi.ac.cy}}

\abstract{We compare several methods for computing disconnected fermion loops contributing
to nucleon three-point functions. The comparison is carried out using one ensemble of
$N_f=2+1+1$ twisted mass fermions with pion mass of 373~MeV. The complete set of operators
up to one-derivative are examined by developing optimized code for mutli-GPUs. Simple
guidelines are given as to the preferable method for each class of operators.}

\FullConference{$31^{st}$ International Symposium on Lattice Field Theory LATTICE 2013\\
July 29 – August 3, 2013\\
Mainz, Germany}

\begin{document}

\section{Introduction}

The evaluation of disconnected quark loops is of paramount importance for the computation
of flavor singlet quantities, but, on the lattice, this requires the calculation of all-
time-slice-to-all propagators, which cannot be carried out by inverting the Dirac matrix at
all lattice points, so stochastic methods are tranditionally used to estimate the inverse
matrix. Provided the number of stochastic noise vectors $N_r$ needed are much less than the
number of lattice points, then this method can be applied efficiently. To reduce the
stochastic noise for operators requiring a large number of $N_r$, we applied the truncated
solver method~\cite{TSM}; however, disconnected fermion loops are proned to gauge noise.
Therefore, one needs a large number of statistics as well as other noise reduction techniques.
In this work we analyze the efficiency of several variance reduction methods for twisted mass
fermions, implemented on GPUs.

\section{Stochastic methods}
\label{sec:disc_methods}
A direct computation of the inverse of the fermionic matrix, whose size ranges from $\sim 10^7$
to $\sim 10^9$ for the largest volumes considered nowadays, is not feasible with our current
computer power. Nonetheless, we can calculate an unbiased stochastic estimate of the inverse by
generating a set of $N_r$ random sources $\left|\eta_j\right\rangle$, filling each component
with $\mathbb{Z}_N$ noise with the following properties:

\begin{minipage}{0.45\linewidth} 
\begin{equation} 
\frac{1}{N}\sum_{j=1}^{N_r}\left|\eta_j\right\rangle = O\left(\frac{1}{\sqrt{N_r}}\right),
\label{eq1}
\end{equation} 
\end{minipage} 
\hspace{0.04\linewidth} 
\begin{minipage}{0.45\linewidth} 
\begin{equation} 
\frac{1}{N_r}\sum^{N_r}_{j=1}\left|\eta_j\right\rangle\left\langle\eta_j\right| = \mathbb{I} +
{\cal O}\left(\frac{1}{\sqrt{N_r}}\right).
\label{eq2}
\end{equation} 
\end{minipage}

The first property ensures that our estimate of the propagator is unbiased. The second one
allows us to reconstruct the inverse matrix by solving for $\left|s_r\right\rangle$ in
\begin{eqnarray}
M\left|s_r\right\rangle = \left|\eta_r\right\rangle\quad\longrightarrow
\label{etaToS}
&\quad M_E^{-1}:=\frac{1}{N_r}\sum_{r=1}^{N_r}\left|s_r\right\rangle
\left\langle\eta_r\right|\approx M^{-1}.
\label{estiM}
\end{eqnarray}
The error in our estimate decreases as ${\cal O}\left(1/\sqrt{N_r}\right)$. $\mathbb{Z}_4$ noise sources
were used for this work.

\subsection{The Truncated Solver Method\label{secTSM}}

The Truncated Solver Method (TSM)~\cite{TSM} is a way to increase $N_r$ at a reduced computational cost.
Instead of solving to high precision Eq.~(\ref{estiM}), we can obtain a low precision (LP) estimate 
where the inverter, a CG solver in this work, is truncated. The truncation criterion can
be a large value of the residual $\hat{r}$, or a fixed number of iterations. This way we
can increase the number of stochastic sources $N_{\rm LP}$ at a very small cost. However,
the LP sources will produce a biased estimate of $M^{-1}_E$. This can be corrected by
including  a few  high precision inversions together with the low precision ones, and
calculating the difference as follows

\begin{equation}
M_{E_{TSM}}:=\underbrace{\frac{1}{N_{\rm HP}}\sum_{j=1}^{N_{\rm HP}}
\left[\left|s_j\right\rangle_{HP} - \left|s_j\right\rangle_{LP}\right]
\left\langle\eta_j\right|}_{Correction} + \underbrace{\frac{1}{N_{\rm LP}}
\sum_{j=N_{\rm HP}+1}^{N_{\rm HP}+N_{\rm LP}}\left|s_j\right\rangle_{LP}
\left\langle\eta_j\right|}_{Biased\quad estimate},
\label{estiTSM}
\end{equation}
which requires $N_{\rm HP}$ high precision inversions and $N_{\rm HP}+N_{\rm LP}$ low
precision inversions. If enough sources are used for the correction, the error of this
improved estimator scales as $\propto 1/\sqrt{N_{\rm LP}}$.

\begin{figure}[h!]
   \begin{center}
      \begin{minipage}{0.48\linewidth}
	\includegraphics[width=\linewidth,angle=0]{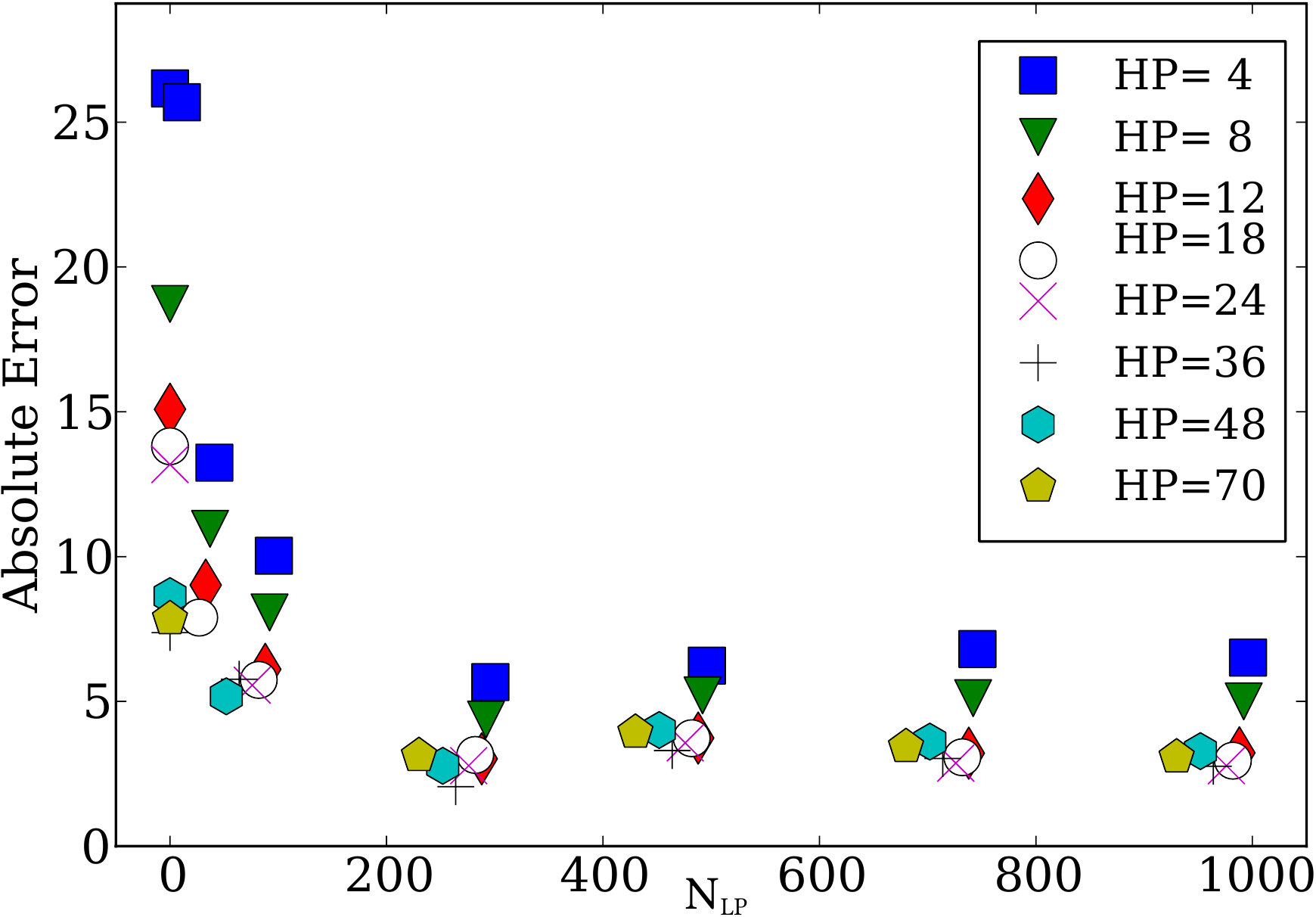}
      \end{minipage}
       \begin{minipage}{0.48\linewidth}
        \includegraphics[width=\linewidth,angle=0]{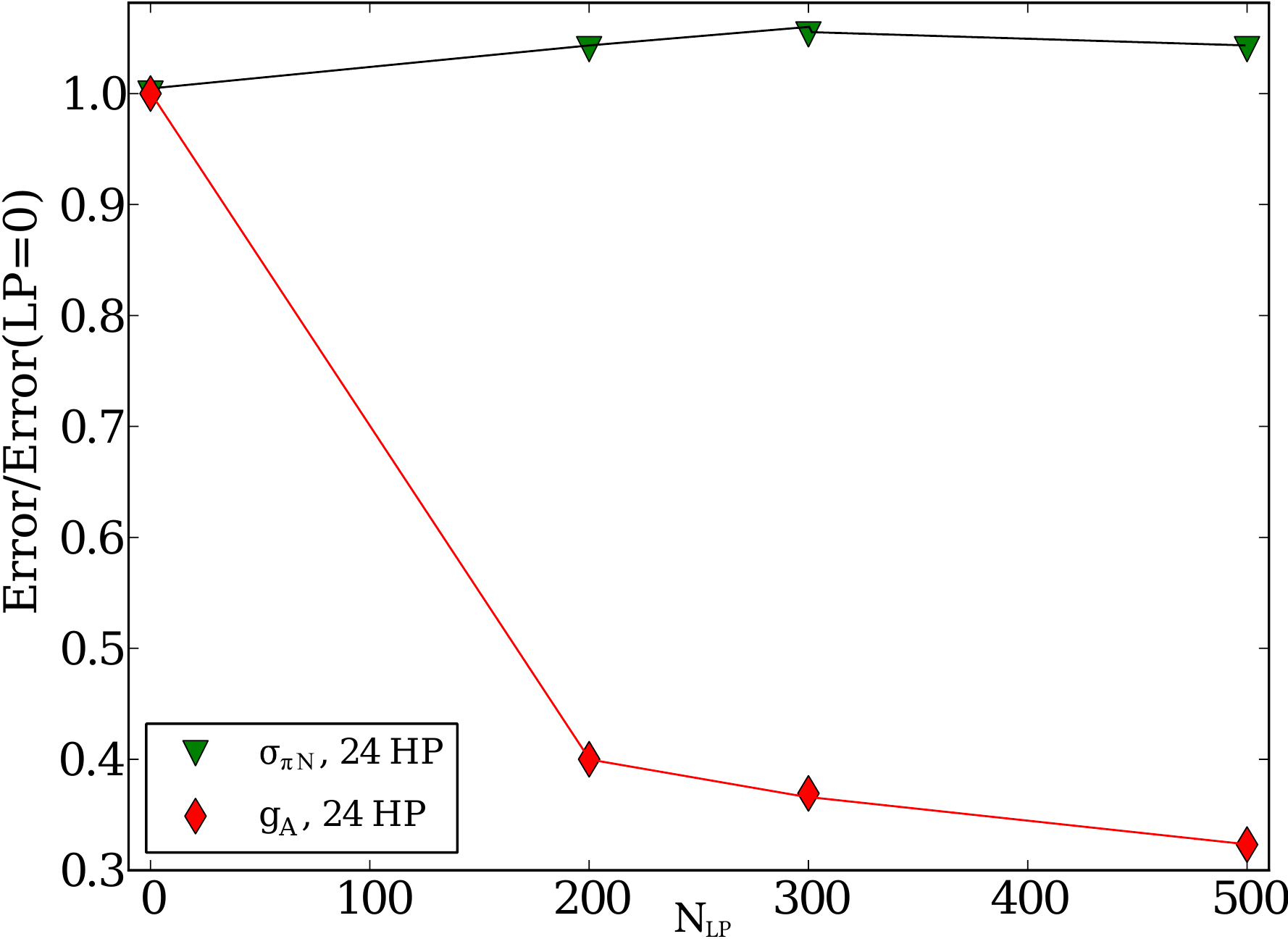} 
      \end{minipage}
      \caption{\footnotesize Left:
      Results on the error of the operator $i\bar\psi\gamma_3 D_3\psi$ versus $N_{\rm LP}$ for 50 measurements Right: Data
      for $\sigma_{\pi N}$ (black line) and $g_A$ (red line) for  56400 measurements. The time of the operator insertion
      $t_{\rm ins} = 8$ and the sink time $t_{\rm s} = 16$ with th source taken at time zero.
      \label{TSMPerfPlot}}
   \end{center}
\end{figure}

In order to achieve optimal performance of the TSM we must tune several parameters. The first
issue is to determine the truncation criterion for the low precision inversions. In our case,
we choose as stopping condition a fixed value for the residual $|\hat{r}|_{\rm LP}\sim10^{-2}$.
The second parameter is the number of $N_{\rm HP}$ required to correct the bias introduced when
using $N_{\rm LP}$ low precision vectors. To fix these parameters, we performed empirical test
upon a reduced set of configurations. As shown in Fig.~\ref{TSMPerfPlot}, different
insertions behave in different ways and might require different tuning.

\vspace{-0.5cm}
\subsection{The one-end trick}

The twisted mass fermion formulation allows the use of the \emph{one-end trick}~
\cite{vvTrick1,vvTrick2} to reduce the variance of the stochastic estimate of disconnected
diagrams. If the operator $X$ has an isovector-flavor structure in the
twisted basis, then one can use
the identity $M^{-1}_u - M^{-1}_d = -2i\mu aM_d^{-1}\gamma_5M_u^{-1}$ to write
 the loop as
\begin{equation}
\frac{2i\mu a}{N_r}\sum_{r=1}^{N_r} \left\langle s^\dagger_r \gamma_5 X s_r\right\rangle
= \textrm{Tr}\left(M_u^{-1}X\right) - \textrm{Tr}\left(M_d^{-1}X\right)
+ O\left(\frac{1}{\sqrt{N_r}}\right).
\label{loopVv}
\end{equation}
With this substitution the fluctuations are reduced by the $\mu$ factor, which should be
small in a reasonable simulation. Also, there is an implicit sum of $V$ terms in Eq.~(\ref{loopVv}),
which improves the signal to noise ratio from $1/\sqrt{V}$ to $V/\sqrt{V^2}$. Unfortunately this technique
can only be applied to operators having a $\tau_3$ flavor matrix in the twisted basis. For operators which do not
have a $\tau_3$ flavor matrix in the twisted basis, we can use instead
\begin{equation}
\frac{2}{N_r}\sum_{r=1}^{N_r} \left\langle s^\dagger_r \gamma_5 X\gamma_5 D_W s_r\right\rangle 
= \textrm{Tr}\left(M_u^{-1}X\right) + \textrm{Tr}\left(M_d^{-1}X\right)
+ O\left(\frac{1}{\sqrt{N_r}}\right).
\label{loopStD}
\end{equation}
However, this generalization lacks the $\mu$-suppression factor, we thus expect
that for this class of operators the fluctuations to be larger. 
Because of the volume sum introduced by our identities, the sources must have entries on
all sites, which in turn means that we compute the fermion loop at all insertion times
simultaneously.

\subsection{Time-dilution}

A well-known variance reduction technique is time-dilution \cite{tDil}, i.e. instead of
filling up all the entries of the source vector, we decompose the whole space $\mathcal{R}=
V\oplus$color$\oplus$spin in $S$ smaller subspaces $\mathcal{R}=\sum_{t=1}^S\mathcal{R}_i$,
one per time-slice, and we define our noise sources at each time-slice. As the noise on one
time-slice contributes to the signal only on this time-slice, but to the noise
on all the other time-slices, time-dilution should reduce the stochastic error.
In addition, one can apply the coherent source method~\cite{Coh} using noise vectors with entries
on several time slices, as long as these time-slices are far enough from each other, so
that they don't interfere with each other.

Time-dilution has a disadvantage for  operators involving a time derivative, since additional inversions at time-slices $t-a$ and
$t+a$ are needed, tripling the computer cost. Therefor time dilution is
benchmarked only for ultra-local
current insertions.

\subsection{Hopping Parameter Expansion}

Another technique to reduce the variance is the \emph{Hopping Parameter Expansion} (HPE)~
\cite{HPE}. The idea is to expand the inverse of the fermionic matrix in terms of the
hopping parameter $\kappa$ as:

\begin{equation}
M_u^{-1} = B - BHB + \left(BH\right)^2B - \left(BH\right)^3B + \left(BH\right)^4M_u^{-1},
\label{HPE}
\end{equation}
where $B=\left(1+i2\kappa\mu a\gamma_5\right)^{-1}$ and $H=2\kappa \slashed{D}$, with $H$
the hopping term. The first four terms in this expansion can be computed exactly, while
the fifth term is calculated stochastically via
\begin{equation}
\frac{1}{N_r}\sum_{r=1}^{N_r} \left[X\left(BH\right)^4 s_r \eta^\dagger_r\right] =
\textrm{Tr}\left[X\left(BH\right)^4 M_u^{-1}\right] + O\left(\frac{1}{\sqrt{N_r}}\right).
\label{loopHPE1}
\end{equation}
All terms involved in Eq.~(\ref{HPE}) are computed in advance and don't depend on the gauge configuration for local
operators, so they do not incur a serious computational overhead. If one expands
the inverse $M_u^{-1}$ to a higher order, then one would have to deal with terms like
$\left(BH\right)^4B$, involving the plaquette or, for high enough orders, with
$\left(BH\right)^{2n}B$, involving $2n$-link structures.

\section{Simulation details}
In order to compare these methods with each other, we consider an ensemble of $N_f=2+1+1$
twisted mass fermions with 4697 gauge configurations. The pion mass is $m_\pi=373$~MeV, with the
the strange and charm quark masses fixed to approximately their physical values. The lattice
spacing of the ensemble is $a=0.082(1)$ determined form the nucleon mass, and the volume $32^3
\times 64$, giving $m_{\pi}L\sim 5$. For the disconnected diagrams we use of the branch
\emph{discLoop} of the QUDA library~\cite{Clark:2009wm,Babich:2011np}. Details on the implementation
can be found in Ref.~\cite{Yo2}.

\section{Comparison of different methods}

\begin{figure*}[h!]
  \begin{center}
    \begin{minipage}{0.32\linewidth}
      \includegraphics[width=\linewidth,angle=0]{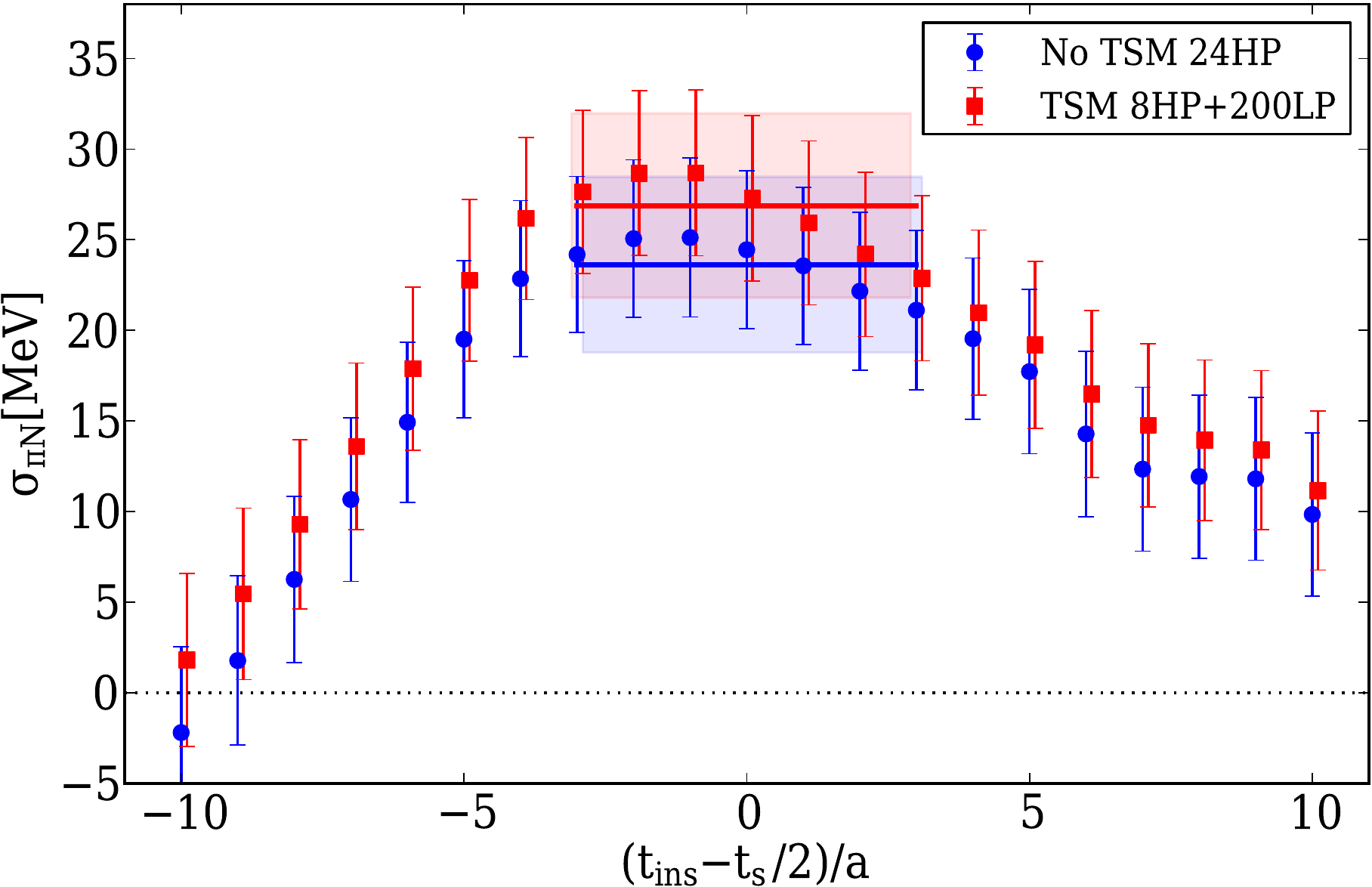}
    \end{minipage}
    \begin{minipage}{0.32\linewidth}
      \includegraphics[width=\linewidth,angle=0]{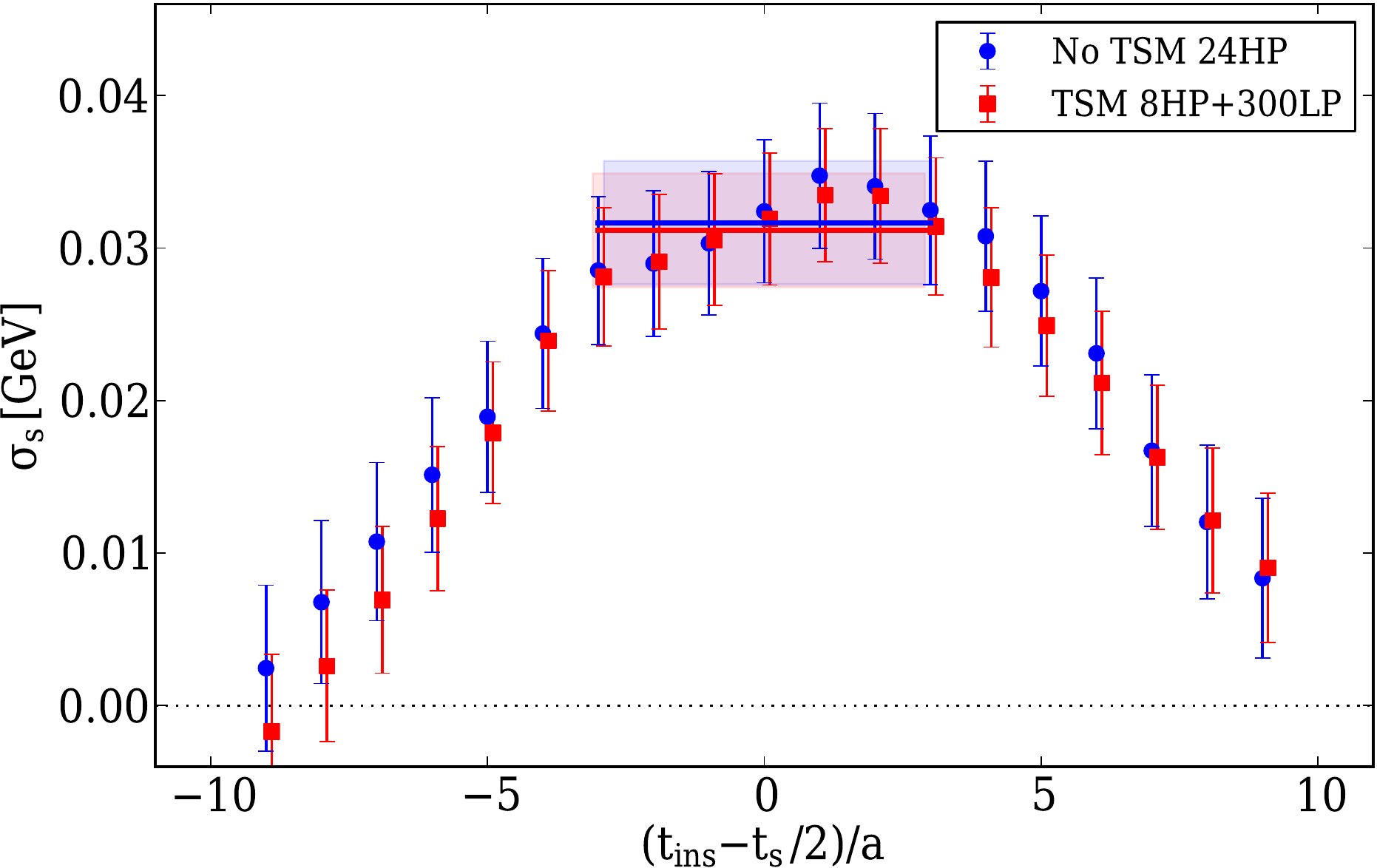} 
    \end{minipage}
    \begin{minipage}{0.32\linewidth}
      \includegraphics[width=\linewidth,angle=0]{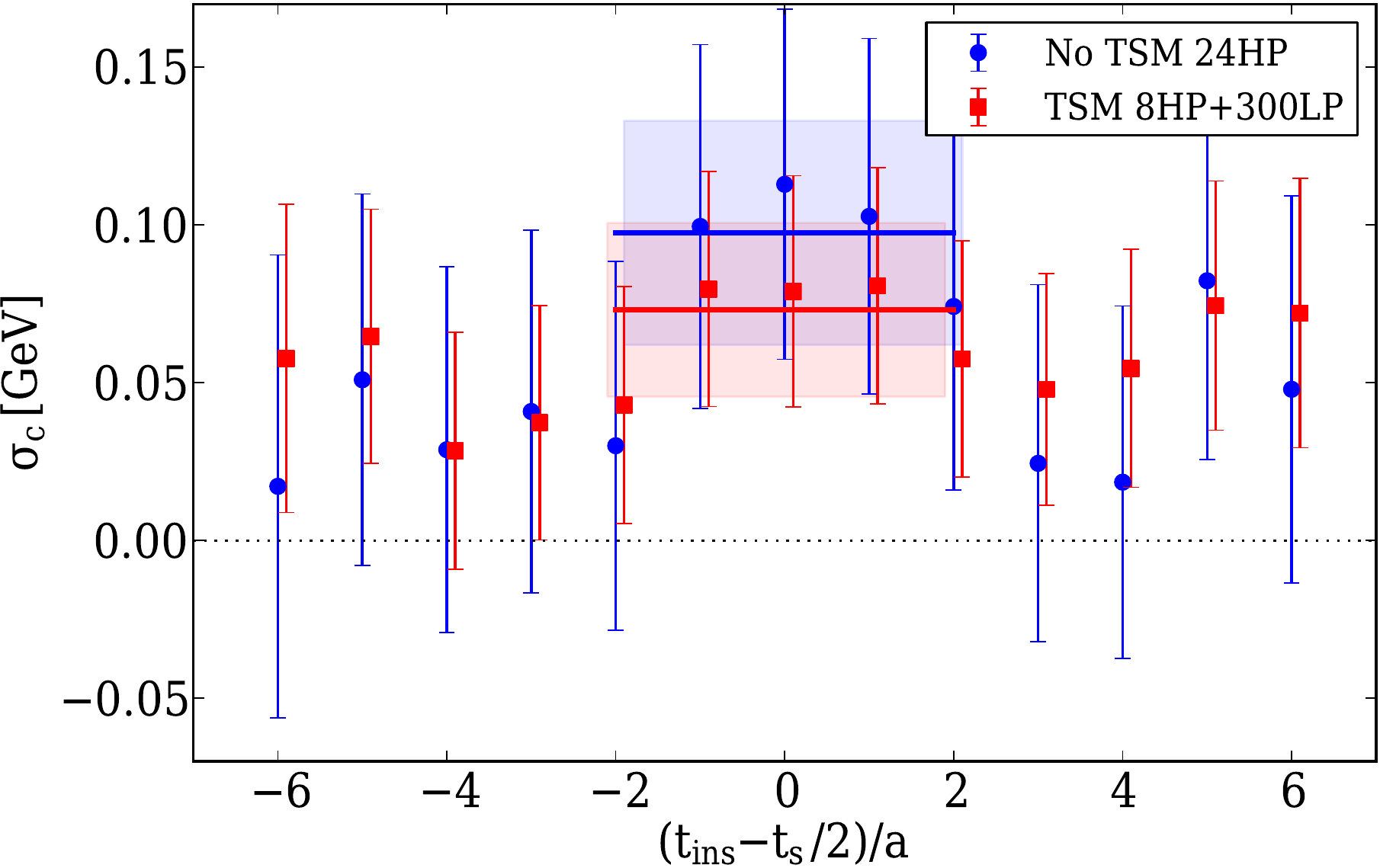} 
    \end{minipage}
        \caption{\footnotesize Comparison of the one-end trick with and without TSM for the disconnected
        contribution to $\sigma_{\pi N}$ (left, 56400 measurements), $\sigma_{s}$ (center,
        58560 measurements) and $\sigma_{c}$ (right, 58560 measurements).
        \label{vsTSMPlots}}
  \end{center}
\end{figure*}

\begin{figure*}[h!]
  \begin{center}
    \begin{minipage}{0.32\linewidth}
      \includegraphics[width=\linewidth,angle=0]{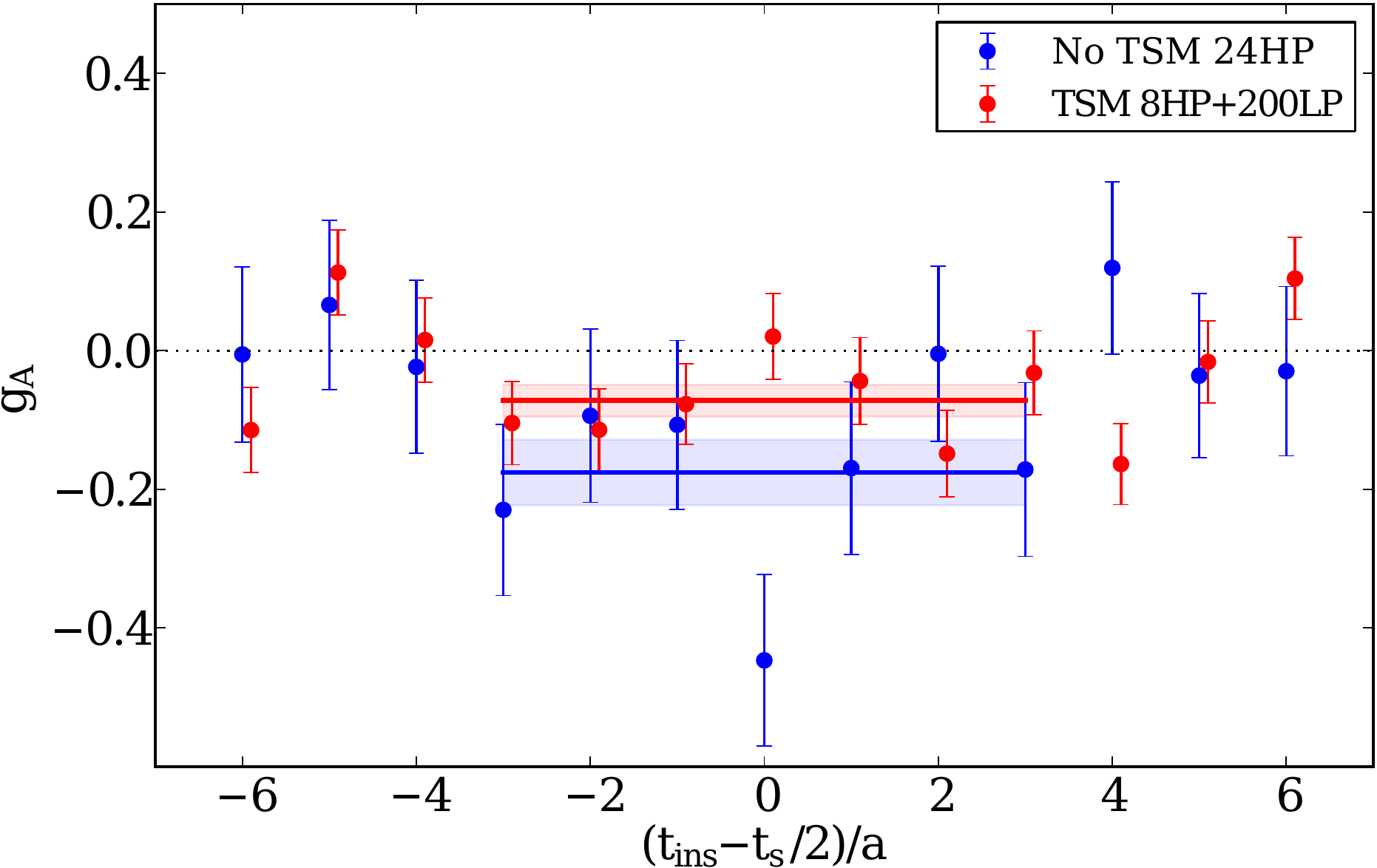}
    \end{minipage}
    \begin{minipage}{0.32\linewidth}
      \includegraphics[width=\linewidth,angle=0]{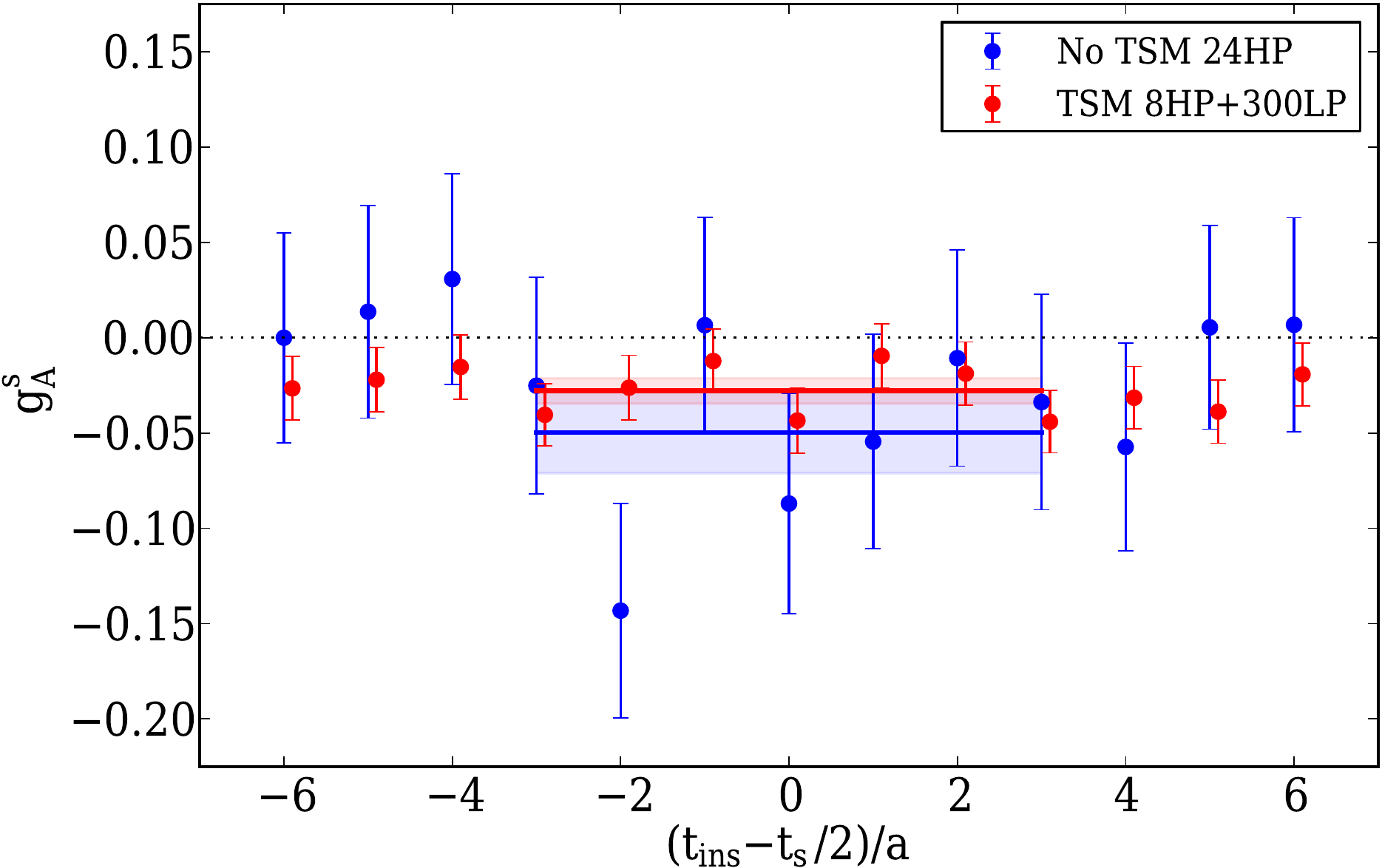} 
    \end{minipage}
    \begin{minipage}{0.32\linewidth}
      \includegraphics[width=\linewidth,angle=0]{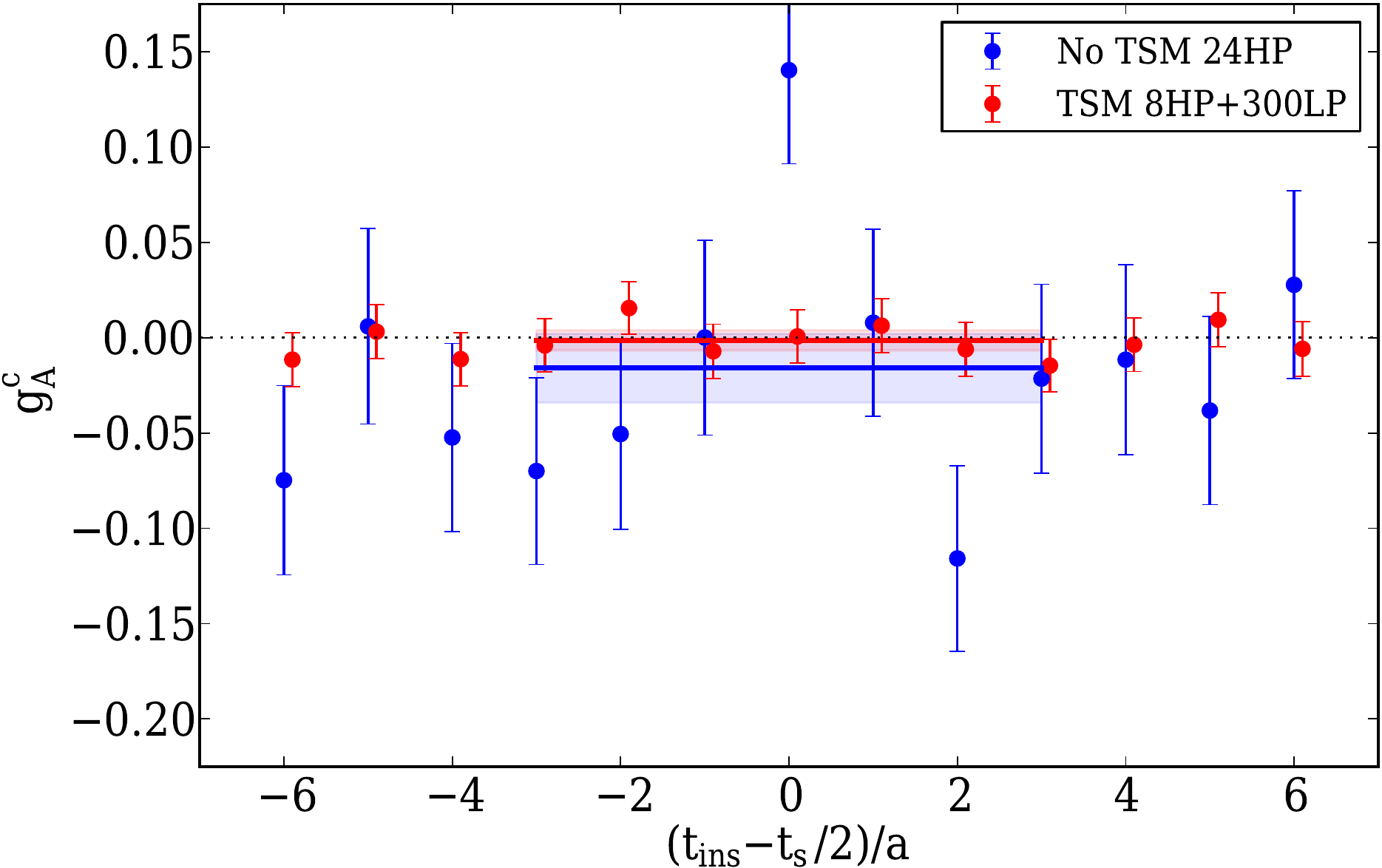} 
    \end{minipage}
    \caption{\footnotesize \label{gAvsTSMPlots} Comparison of the one-end trick with and without TSM for the disconnected
    contribution to isoscalar $g_A$ (left), $g_A^s$ (center) and $g_A^c$ (right). Same
    statistics as in the previous figure.}
  \end{center}
\end{figure*}

\noindent
{\bf Efficiency of TSM:}
In Figs.~\ref{vsTSMPlots} and \ref{gAvsTSMPlots} we show the nucleon $\sigma$-term, for
which the application of the one-end trick brings the $\mu$-noise suppression factor, and
$g_A$, for which it does not, and that is therefore expected to be a more demanding quantity
to compute. We show the disconnected contributions of the light sector, the strange and
charm quark to both of these quantities.

\begin{table}[h!]
\footnotesize
\begin{center}
\begin{tabular}{|c|c|c|c|c|c|}
\hline
				& Light sector	& Strange sector & Charm sector	\\
\hline
$R_{E}$ for $\sigma$-term	&   1.05	&	 0.91    &	0.67	\\
$R_{E}$ for $g_A$		&   0.48	&	 0.30    &	0.28	\\
$R_C$				&   0.66	&	 1.09    &	4.80	\\
$R_CR^2_{E}$ for $\sigma$-term	&   0.73	&	 0.90    &	2.15	\\
$R_C R^2_{E}$ for $g_A$		&   0.15	&	 0.098   &	0.38	\\
\hline 
\end{tabular}
\caption{\footnotesize In the first column we give  $R_E$ and $R_C$ as well as the  quantity
$R_C R^2_E$, which if less than one indicates an advantage of the TSM.\label{tab:TSM}}
\end{center}
\normalsize
\end{table}
In Table~\ref{tab:TSM} we compare the efficiency of TSM by giving the ratio $R_E$ of the
error when using TSM to that without TSM and the ratio $R_C$ of the computer cost with
TSM to the cost without TSM. We also give $R_C R^2_E$, which measures the ratio of efficiencies
independently of the statistics and the error, therefore a value less than one indicates that
the TSM is favorable. The one-end trick is implemented in all cases. For the light quark mass, the
TSM is more efficient for both observables as the product $R_C R^2_E$ indicates. As the
quark mass increases, the advantage of using the TSM is generally reduced. For the charm quark
loops contributing to the $\sigma$-term the TSM ceases to be advantageous whereas for $g_A$ the TSM
is still useful in all range of masses.

\begin{figure*}[h!]
  \begin{center}
    \begin{minipage}{0.235\linewidth}
      \includegraphics[width=\linewidth,angle=0]{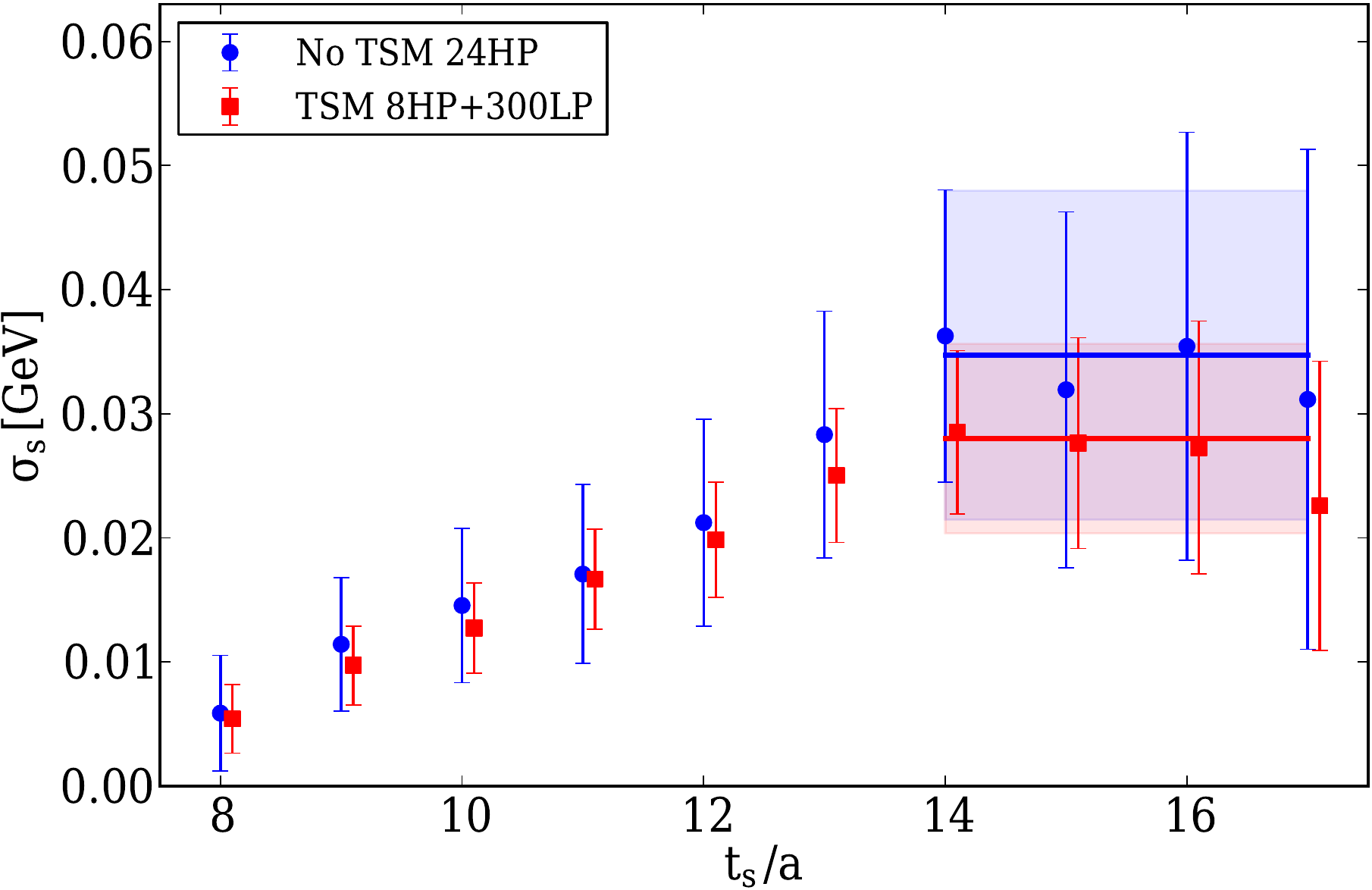}
    \end{minipage}
    \begin{minipage}{0.235\linewidth}
      \includegraphics[width=\linewidth,angle=0]{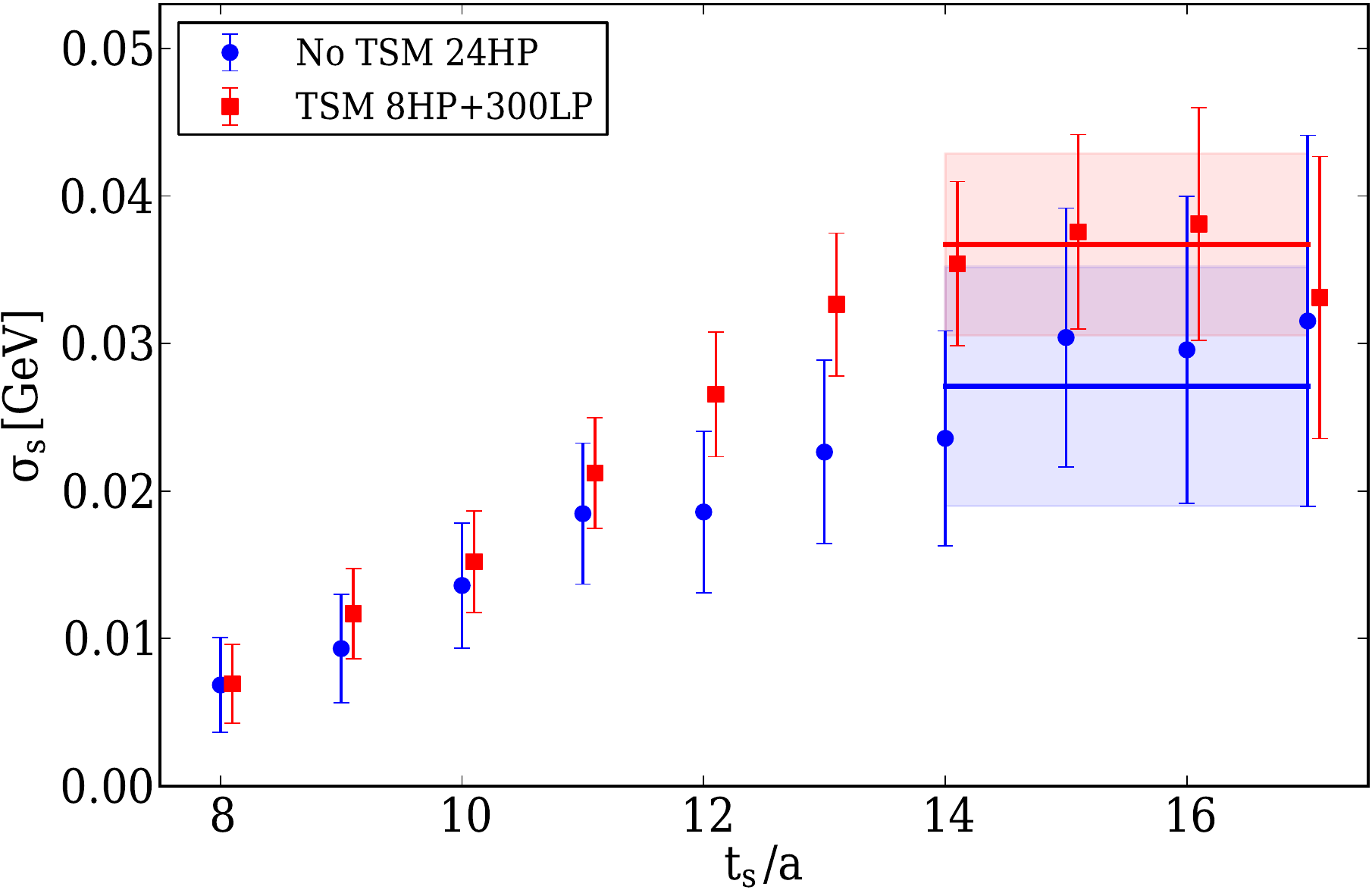} 
    \end{minipage}
    \begin{minipage}{0.235\linewidth}
      \includegraphics[width=\linewidth,angle=0]{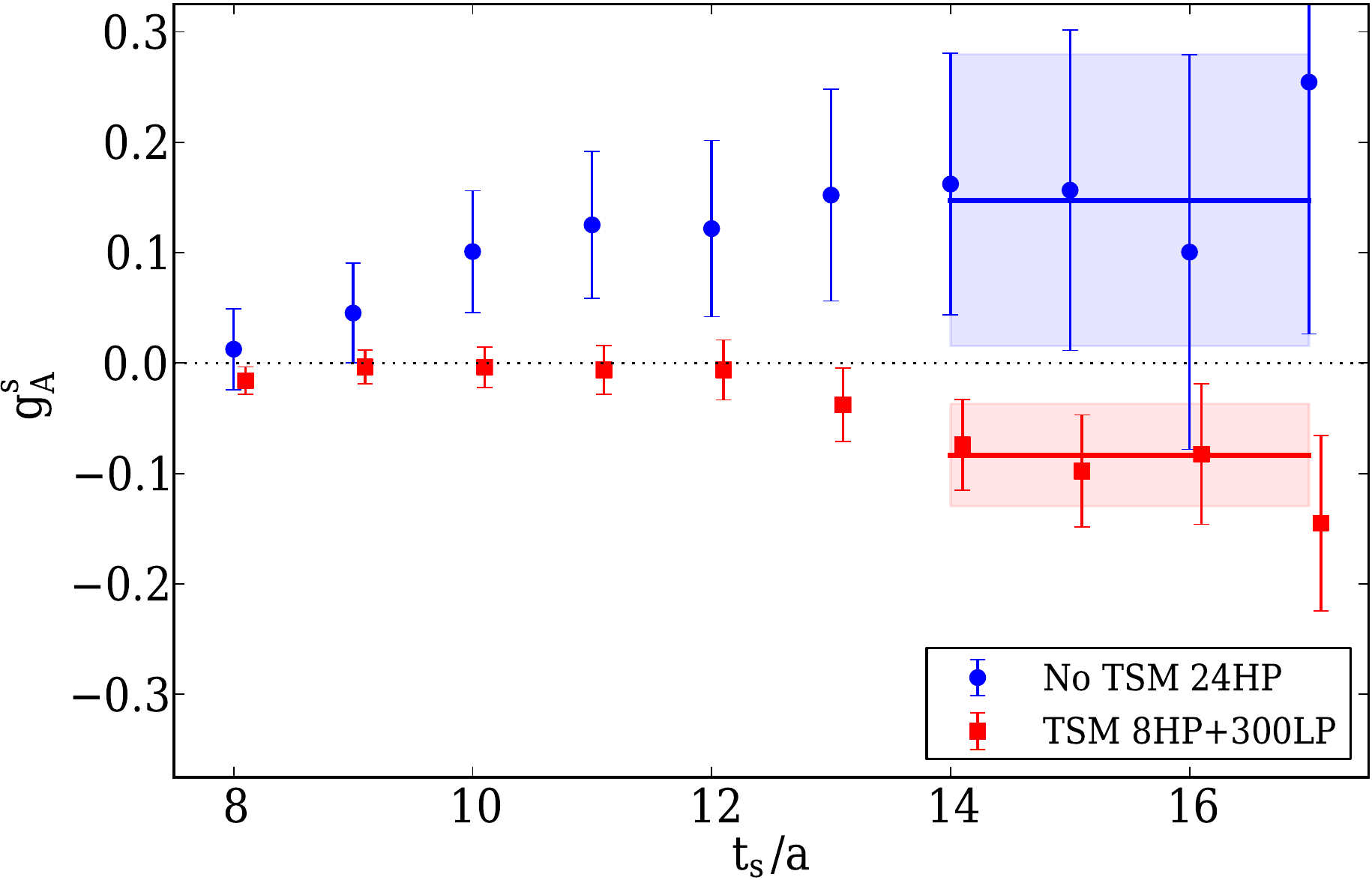}
    \end{minipage}
    \begin{minipage}{0.235\linewidth}
      \includegraphics[width=\linewidth,angle=0]{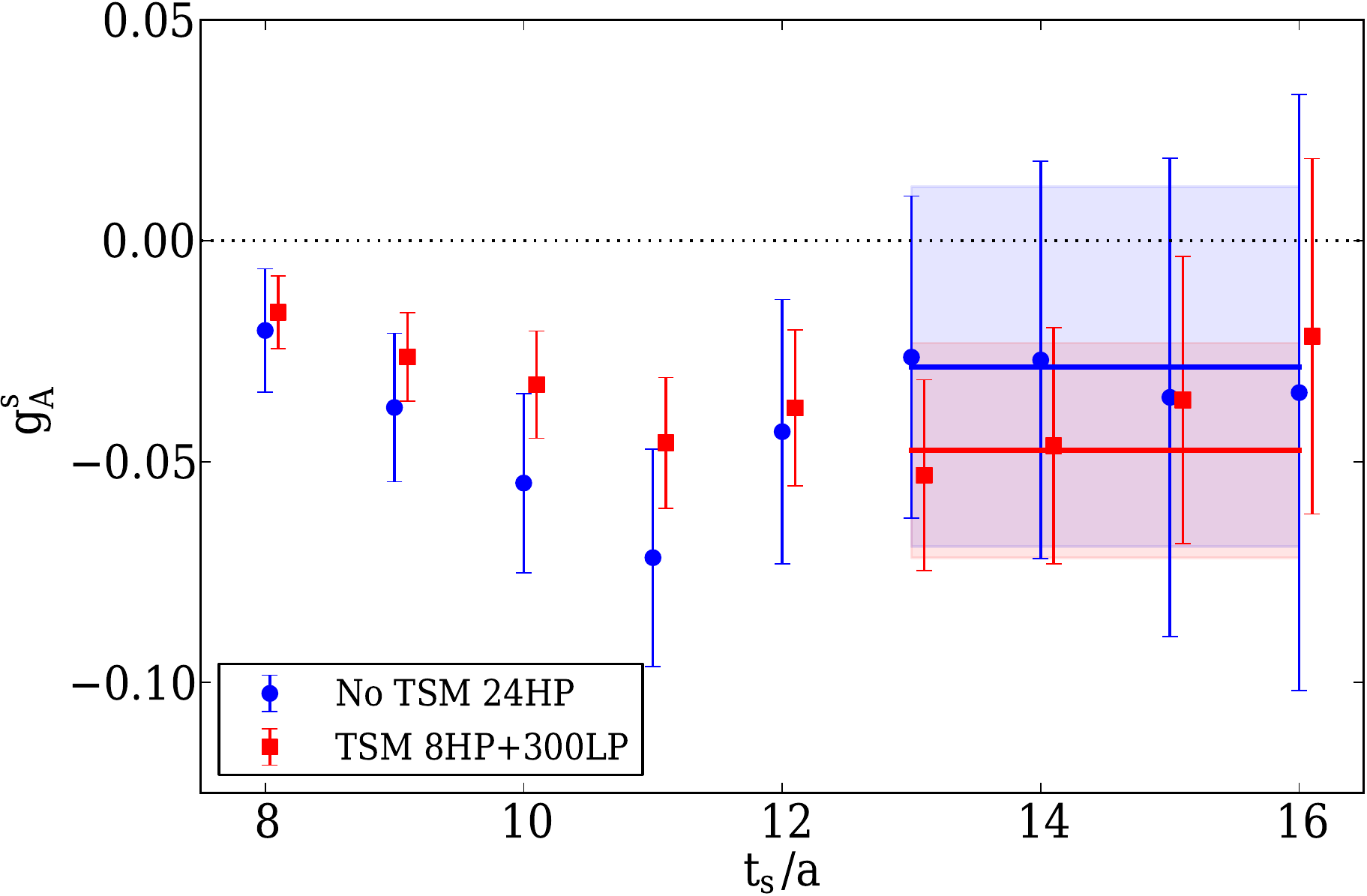} 
    \end{minipage}
        \caption{\footnotesize Comparison of time-dilution plus HPE, with and without the TSM, for the case of
        $\sigma_s$ (first and second to the left) and $g_A^s$ (first and second to the right). The operator insertion is
        $t_{\rm ins} = 8a$ and the number of measurements 18628.}
        \label{tDilSigmavsTSMPlots}
  \end{center}
\end{figure*}

For the case of strange quark loops, we also examine the efficiency of the TSM 
with respect to time-dilution, as well as whether including the HPE gives any
additional benefit. The performance in this case can be assessed easily since
the computational cost is roughly the same.
As shown in Fig.~\ref{tDilSigmavsTSMPlots}, the TSM always reduces the error,
and including HPE is a must, for it comes at virtually no cost, and nearly halves
the error. However, we expect the HPE to perform worse (better) as we decrease
(increase) the quark mass.

A way to measure the efficiency of the TSM is the ratio $R_{\rm HP/LP}$, which is the
number of LP inversions and source contractions one can compute using the time required
for a HP inversion with the associated contractions. Thus, the value of this ratio depends
not only on the time required for the inversions, but also includes time needed to perform
all the contractions to obtain the loops. In Table~\ref{compRatios} we give the ratio $R_{\rm HP/LP}$
for the different methods and quark masses. A large value for this ratio means that
the TSM is advantageous. For the light sector we find a big benefit since the inversions
are much more time consuming  than the contractions. For the charm sector the time needed for
contractions and for a HP inversion are similar, and the TSM brings no benefit.

\begin{table}[h!]
\footnotesize
\begin{center}
\begin{tabular}{|c|c|c|c|c|c|}
\hline
   Method   	    & Quark sector & $R_{\rm HP/LP}^{Local}$ & $R_{\rm HP/LP}^{One-Deriv.}$ \\
\hline
One-end trick	    &    Light     &       $\sim26.7$        & $\sim10$ \\
One-end trick	    &    Strange   &       $\sim16.9$        & $\sim5.8$ \\
One-end trick	    &    Charm     &       $\sim 2.9$        & $\sim1.4$ \\
Time-dilution       &    Strange   &       $\sim20.7$        & --- \\
Time-dilution + HPE &    Strange   &       $\sim19.1$        & --- \\
\hline 
\end{tabular}
\caption{\footnotesize The $R_{\rm HP/LP}$ ratio for the different methods for light, strange and charm quark
loops. In the third column the ratio for all ultra-local operators is given and in the fourth
column all one-derivative operators are also included in $R_{\rm HP/LP}$.}
\label{compRatios}
\end{center}
\normalsize
\end{table}

\noindent
{\bf Time-dilution plus HPE vs the one-end trick:}
Besides comparing the advantages of the TSM, it would be interesting to compare the one-end trick
to time-dilution with HPE. 

\begin{figure*}[h!]
  \begin{center}
    \begin{minipage}{0.235\linewidth}
      \includegraphics[width=\linewidth,angle=0]{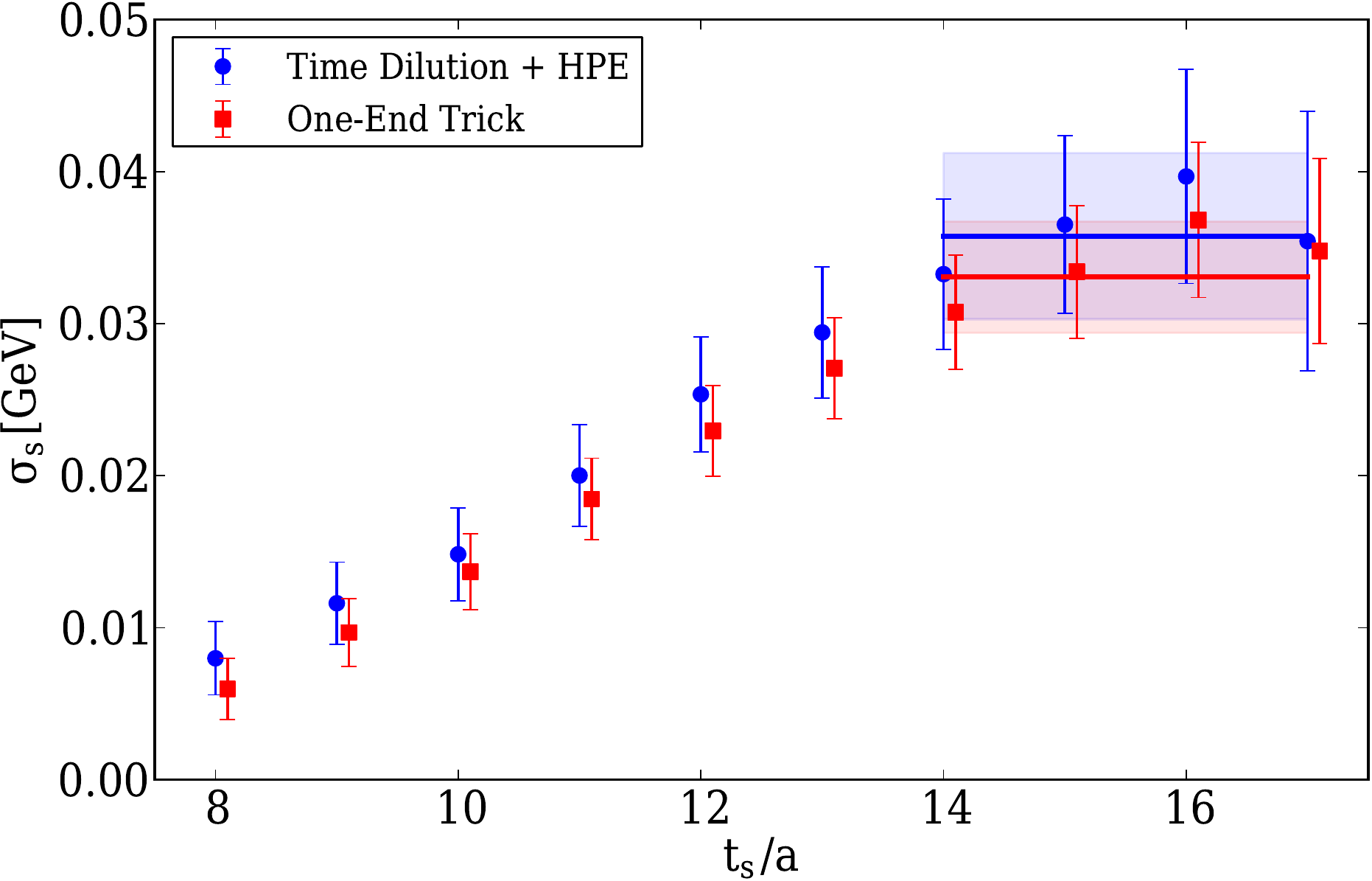}
    \end{minipage}
    \begin{minipage}{0.235\linewidth}
      \includegraphics[width=\linewidth,angle=0]{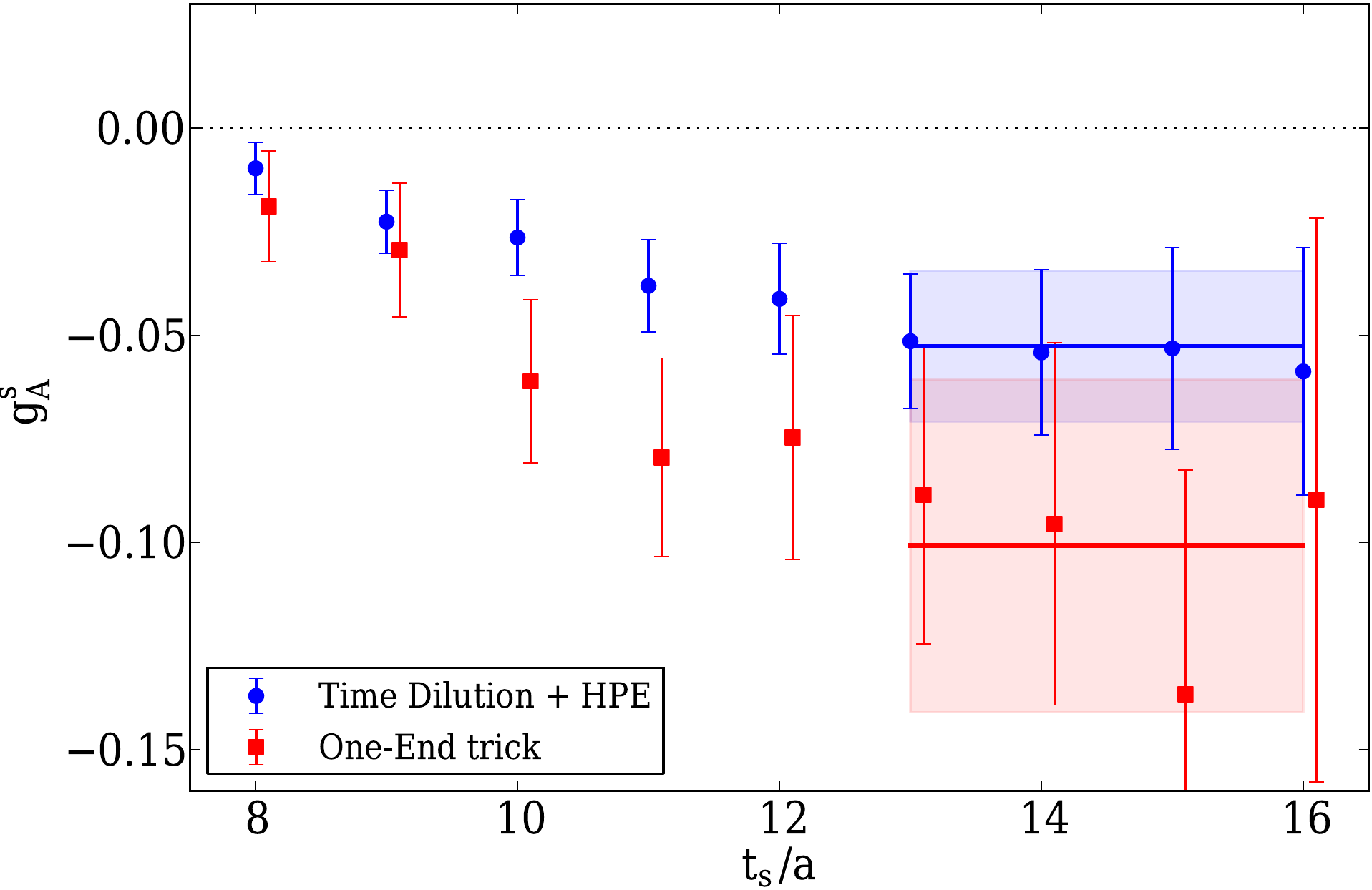} 
    \end{minipage}
    \begin{minipage}{0.235\linewidth}
      	\includegraphics[width=\linewidth,angle=0]{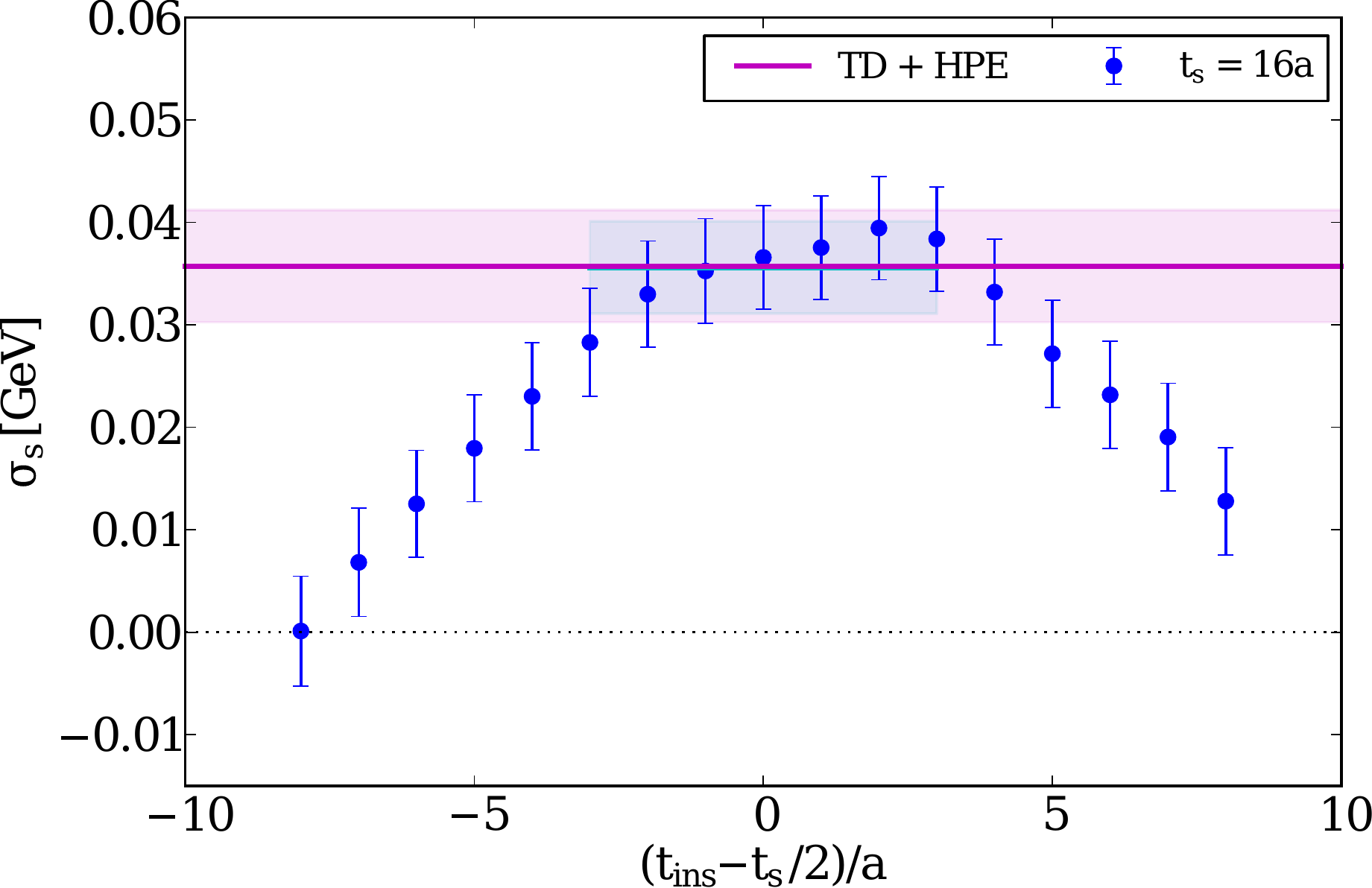}
    \end{minipage}
    \begin{minipage}{0.235\linewidth}
      \includegraphics[width=\linewidth,angle=0]{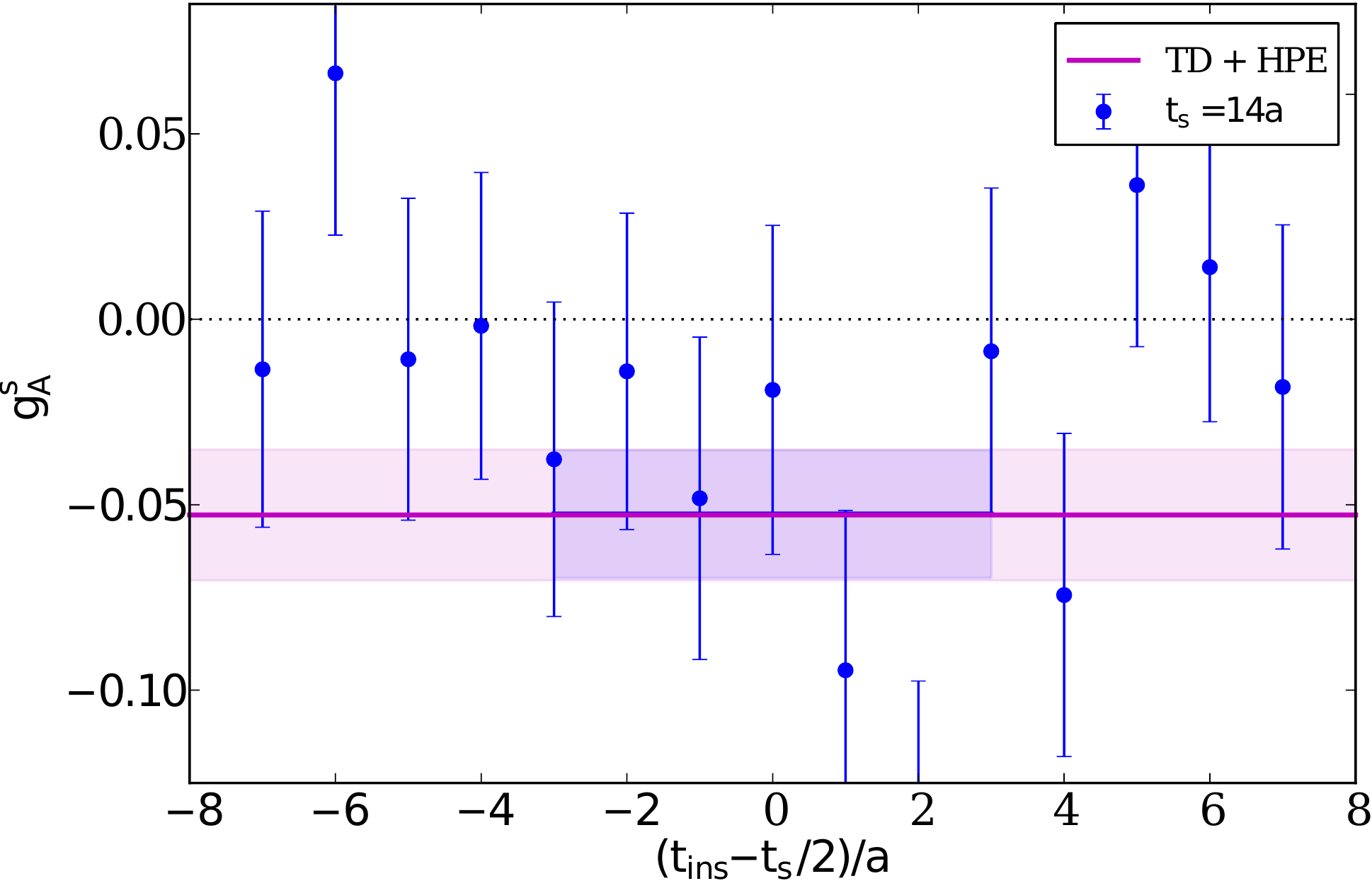} 
    \end{minipage}
        \caption{\footnotesize Comparison of results when using the one-end trick plus TSM ($N_{\rm HP}=24$ and $N_{\rm LP}=300$)
	to using time-dilution plus HPE plus TSM ($N_{\rm HP}=24$ and $N_{\rm LP}=300$), same statistics, for
	$\sigma_s$ (leftmost) and $g_A^s$ (second from the left). The number of measurement is 18628 and the
	current method was used with $t_{\rm ins}=8a$. In the first right and rightmost panels we show the same 
	quantities, but computed using the fixed sink method for the one-end trick. The purple band is the
	value of the plateau when time-dilution is used with the fixed current method.}
        \label{tDilVsVV}
  \end{center}
\end{figure*}
Results are shown in Fig.~\ref{tDilVsVV} for $\sigma_s$ and $g_A^s$. For $\sigma_s$ time-dilution
gives larger errors as compared to the one-end trick for the same statistics, while for $g_A^s$
considerably smaller errors are obtained with time-dilution. Nonetheless, with the one-end trick one
obtains the quark loops at all time-slices, yielding effectively more measurements. This also allows
to vary the insertion time-slide and to fit to a plateau as shown by the blue band in the rightmost
plot of Fig.~\ref{tDilVsVV}. As can be seen, this plateau value has the same error as the one extracted
from fitting the asymptotic behavior of the ratio computed using time-dilution with HPE (purple band).
Therefore the one-end trick, having the advantage of yielding all time-slides, can perform as well as
time-dilution with HPE, also in the case of $g_A^s$.

\section{Conclusions}
The computation of disconnected contributions has become feasible due to improvements in algorithms and
computational power. In this work, we compare several different strategies to calculate disconnected
diagrams by using the GPU-optimized library QUDA on its discLoop branch.

Our comparison shows that the one-end trick with the TSM is the optimal method for the computation of
the light and strange quark loops with an ultra-local  and one-derivative operator insertions, whereas for
the charm quark loops, we prefer time-dilution with the HPE and TSM for ultra-local operators, and the one-end trick
for one-derivative insertions. The last choice is justified by the increase in the number of inversions required to
apply time-dilution.

\section*{Acknowledgments}  
A. Vaquero and K. Jansen are supported by funding from the Cyprus RPF under contract
EPYAN/0506/08 and $\Pi$PO$\Sigma$E$\Lambda$KY$\Sigma$H/EM$\Pi$EIPO$\Sigma$/0311/16 respectively.
This research was in part supported by the Research Executive Agency of the
EU under Grant Agreement number PITN-GA-2009-238353 (ITN STRONGnet) and the infrastructure
project INFRA-2011-1.1.20 number 283286, and the Cyprus RPF under contracts
KY-$\Gamma$A/0310/02 and NEA Y$\Pi$O$\Delta$OMH/$\Sigma$TPATH/0308/31. Computer resources
were provided by Cy-Tera at CaSToRC, Forge at NCSA Illinois (USA), Minotauro at BSC
(Spain), and Jugene Blue Gene/P at the JSC, awarded under the 3rd PRACE call.

\bibliography{ref}

\end{document}